\documentclass[acmsmall]{acmart}
\AtBeginDocument{%
  }

\setcopyright{cc}
\setcctype{by-nc-nd}
\acmDOI{10.1145/3832169}
\acmYear{2026}
\acmJournal{PACMSE}
\acmVolume{3}
\acmNumber{ISSTA}
\acmArticle{ISSTA078}
\acmMonth{10}
\acmSubmissionID{issta26main-p733-p}
\received{2026-01-30}
\received[accepted]{2026-04-16}

\usepackage{xspace}
\usepackage{listings, listings-rust}
\usepackage{color}
\usepackage[table]{xcolor}
\usepackage{pdfpages}
\usepackage{graphicx}
\usepackage{mathtools}
\usepackage{amsmath}
\usepackage{wasysym}
\usepackage{caption}
\usepackage{subcaption}
\usepackage[ruled,vlined,linesnumbered]{algorithm2e}
\usepackage{enumitem}
\usepackage{multirow}
\usepackage{minted}
\usepackage{marvosym}
\usepackage{colortbl}

\newcommand*{\cetus}{\textit{Cetus}\xspace}
\newcommand*{\nemo}{\textit{Nemo}\xspace}

\newcommand{\shaohua}[1]{{#1}}
\newcommand{\wanxu}[1]{{#1}}

\newcommand{\zq}[1]{#1}
\newcommand{\issta}[1]{#1}

\newcommand{\rev}[1]{\textcolor{black}{#1}}

\newcommand{\para}[1]{\medskip\noindent\textbf{#1}\xspace}
\newcommand*{\tool}{{\textsc{Belobog}}\xspace}
\newcommand*{\ityfuzz}{{\textsc{ItyFuzz}}\xspace}
\newcommand*{\suifuzzer}{{\textsc{Sui-Fuzzer}}\xspace}
\newcommand*{\movescan}{{\textsc{MoveScan}}\xspace}

\newcommand{\eg}{\hbox{\emph{e.g.}}\xspace}
\newcommand{\ie}{\hbox{\emph{i.e.}}\xspace}

\makeatletter
\renewcommand{\@fnsymbol}[1]{%
  \ensuremath{%
    \ifcase#1\or *\or \text{\Letter}\or \dagger\or \ddagger\or
    \mathsection\or \mathparagraph\or \|\or **\or \dagger\dagger\or
    \ddagger\ddagger\else\@ctrerr\fi}}
\makeatother

\definecolor{codegreen}{rgb}{0,0.6,0}
\definecolor{codegray}{rgb}{0.8,0.5,0.5}
\definecolor{codepurple}{rgb}{0.58,0,0.82}
\definecolor{backcolour}{rgb}{0.95,0.95,0.92}
\lstdefinestyle{mystyle}{
    backgroundcolor=\color{backcolour},   
    commentstyle=\color{codegreen},
    keywordstyle=\color{magenta},
    numberstyle=\tiny\color{codegray},
    stringstyle=\color{codepurple},
    basicstyle=\ttfamily\footnotesize,
    breakatwhitespace=false,         
    breaklines=true,                 
    captionpos=b,                    
    keepspaces=true,                 
    numbers=left,                    
    numbersep=5pt,                  
    showspaces=false,                
    showstringspaces=false,
    showtabs=false,                  
    tabsize=2,
    xleftmargin=12pt,
    otherkeywords={function, external, internal, address, uint256}
}
\begin{document}

\title{Belobog: Move Language Fuzzing Framework for Real-World Smart Contracts}


\author{Ziqiao Kong}
\authornotemark[1]
\orcid{0009-0009-4926-4932}
\affiliation{%
  \institution{Nanyang Technological University}
  \city{Singapore}
  \country{Singapore}
}
\email{ziqiao001@e.ntu.edu.sg}

\author{Wanxu Xia}
\authornote{Both authors contributed equally to this research.}
\orcid{0009-0007-4241-5983}
\affiliation{%
  \institution{Beihang University}
  \city{Beijing}
  \country{China}
}
\email{ysiel@buaa.edu.cn}

\author{Zhengwei Li}
\orcid{0009-0008-3803-6347}
\affiliation{%
  \institution{Bitslab}
  \city{Singapore}
  \country{Singapore}
}
\email{robin@bitslab.xyz}

\author{Yi Lu}
\orcid{0009-0000-4238-8881}
\affiliation{%
  \institution{Movebit}
  \city{Singapore}
  \country{Singapore}
}
\email{y@movebit.xyz}

\author{Pan Li}
\orcid{0009-0007-2106-9559}
\affiliation{%
  \institution{Bitslab}
  \city{Singapore}
  \country{Singapore}
}
\email{paul@bitslab.xyz}

\author{Liqun Yang}
\authornotemark[2]
\orcid{0000-0002-5498-3474}
\affiliation{%
  \institution{Beihang University}
  \city{Beijing}
  \country{China}
}
\email{lqyang@buaa.edu.cn}

\author{Yang Liu}
\orcid{0000-0001-7300-9215}
\affiliation{%
  \institution{Nanyang Technological University}
  \city{Singapore}
  \country{Singapore}
}
\email{yangliu@ntu.edu.sg}

\author{Xiapu Luo}
\orcid{0000-0002-9082-3208}
\affiliation{%
  \institution{Hong Kong Polytechnic University}
  \city{Hong Kong}
  \country{Hong Kong}
}
\email{csxluo@comp.polyu.edu.hk}

\author{Shaohua Li}
\authornote{Both authors are corresponding authors.}
\orcid{0000-0001-7556-3615}
\affiliation{%
  \institution{The Chinese University of Hong Kong}
  \city{Hong Kong}
  \country{Hong Kong}
}
\email{shaohuali@cse.cuhk.edu.hk}

\renewcommand{\shortauthors}{Ziqiao Kong et al.}

\begin{abstract}
Move is a resource-oriented programming language designed for secure and verifiable smart contract development and has been widely used in managing billions of digital assets in blockchains, such as Sui and Aptos.
Move features a strong static type system \shaohua{and explicit resource semantics} to enforce safety properties such as the prevention of data races, invalid asset transfers, and entry vulnerabilities.
However, smart contracts written in Move may still contain certain vulnerabilities that are beyond the reach of its type system.
It is thus essential to validate Move smart contracts. Unfortunately, due to its strong type system, existing smart contract fuzzers are ineffective in producing syntactically or semantically valid transactions to test Move smart contracts.

This paper introduces the first fuzzing framework, \tool, for Move smart contracts.
\tool is type-aware and ensures that all generated and mutated transactions are well-typed.
More specifically, for a target Move smart contract, \tool first constructs a \emph{\wanxu{dependency graph}} based on Move's type system, and then generates or mutates a transaction based on the \emph{graph trace} derived from the \emph{\wanxu{dependency graph}}.
In order to overcome the complex checks in Move smart contracts, we further design and implement a concolic executor in \tool.

We evaluated \tool on \zq{109} real-world Move smart contract projects. The experimental results show that \tool is able to detect 100\% critical and \zq{79\%} major vulnerabilities manually audited by human experts.
We further selected two recent notorious incidents in the Move ecosystem, \ie, \cetus and \nemo. \tool successfully reproduced full exploits for both of them, without any prior knowledge.
\issta{Moreover, we applied \tool on three ongoing auditing projects and found 2 critical, 2 major, and 3 medium new vulnerabilities, all acknowledged by the project developers.}

\end{abstract}



\maketitle

\section{Introduction}
Smart contracts have become a core component of modern blockchain systems, enabling decentralized applications to automate interactions.
Since smart contracts have been used to manage numerous digital assets, their correctness and security are critical. Bugs in deployed smart contracts can result in severe consequences, such as financial loss~\cite{zhou2023sok}.
\zq{The most prevailing smart} contract language, Solidity~\cite{solidity}, adopt an easy-to-use language design but expose developers to subtle security pitfalls, \zq{such as reentrancy attacks.} 
These recurring issues highlight the need for languages that provide stronger security guarantees by design.

The Move language emerges as a new smart contract language designed with strong type safety.
Its static type system and resource-oriented semantics enforce key safety properties at both compile and run time, preventing common bugs like reentrancy.
The Move language has been increasingly adopted by mainstream blockchains. For example, \issta{Sui\cite{suimove} and Aptos\cite{aptosmove}, the two prevailing blockchains with 6.4 billion market capitalization and 1.5 million daily active users~\cite {Artemis}, use the Move language for development.}
Despite its strong type system, smart contracts written in Move can still contain logic flaws or other bugs that are beyond the scope of the type system, such as the notorious bugs in \cetus\cite{cetusreport} and \nemo\cite{nemoreport}, \issta{causing millions of dollars of loss}.
Therefore, it is still essential to test and verify a Move smart contract.
Unfortunately, existing smart contract fuzzers are largely designed for the Ethereum Virtual Machine (EVM)\cite{evm} and lack awareness of Move's unique type system.
As our evaluation in Section~\ref{sec:motivation-existing} will show, they all fail to generate enough valid transactions---most transactions are rejected by the \rev{Move Virtual Machine} before reaching meaningful code paths.

\begin{figure*}[t]
\begin{subfigure}{0.6\textwidth}
    \begin{minted}[xleftmargin=1em,numbersep=3pt,fontsize=\footnotesize,linenos,escapeinside=@@]{rust}
module pool;
// This struct type has "drop" and "store" abilities.
public struct VeryAble has drop,store {...}
// This struct type has neither "drop" nor "store" abilities.
public struct Receipt<T> {...}

public fun loan<T>(amount: u64): (Coin<T>, Receipt<T>) { 
    // Function to flash loan.
    ...
} 
public fun repay<T>(coin: Coin<T>, receipt: Receipt<T>) {
    // Function to repay flash loan with coins.
    ...
}
    \end{minted}
    \caption{A constructed Move smart contract with two struct definitions and two function definitions.}
    \label{fig:intro_example}
\end{subfigure}
\hfill
\begin{subfigure}{0.35\textwidth}
    \centering
    \includegraphics[width=0.8\linewidth]{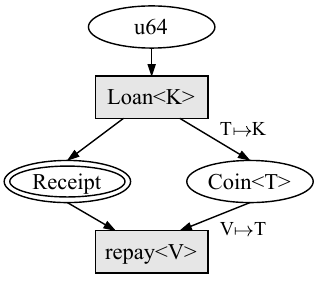}
    \vspace{20pt}
    \caption{The simplified \wanxu{dependency graph} for the constructed Move smart contract.}
    \label{fig:intro_typegraph}
\end{subfigure}
    \caption{A constructed Move smart contract (left) and the built \wanxu{dependency graph} for it (right).}
\end{figure*}

Generating diverse yet valid transactions is crucial for a successful fuzzer design.
We identify three key challenges in fuzzing Move smart contracts, namely \emph{type safety of objects}, \emph{type parameter of functions}, and \emph{type abilities of objects}.
Figure~\ref{fig:intro_example} shows a constructed Move smart contract that exhibits all three challenges.

\noindent\textbf{Challenge 1: Type safety of objects.} 
Move represents states as strongly-typed objects, which results in several strict rules, \ie, (1) objects that are passed as function inputs must match the function's signature, (2) objects can only be created in modules defining them, and (3) an account can only use its owned or public shared objects to interact with Move functions.
Thus, fuzzers cannot forge objects with random bytes but must instead construct them through legitimate function calls or reuse existing ones.
For example, the second argument type of \verb|repay<T>()| in Figure~\ref{fig:intro_example} is \verb|Receipt<T>|, which is defined in the same module as \verb|repay<T>()|. In this module, the only function that can produce objects of type \verb|Receipt<T>| is \verb|loan<T>()|, and thus a valid transaction must call \verb|loan<T>()| first to create a usable object of type \verb|Receipt<T>| before calling \verb|repay<T>()|.

\noindent\textbf{Challenge 2: Type parameter of functions.} Move functions and structs are often generic. Fuzzers must provide valid type instantiations to generate executable transactions, which existing fuzzers fail to do.
For example, both functions \verb|loan<T>()| and \verb|repay<T>()| accept an extra type parameter \verb|T|, which usually specifies the actual currency coin types on blockchains, such as \verb|USDC| and \verb|USDT|.
When constructing a transaction, it is crucial to fill in valid concrete types for all type parameters.

\noindent\textbf{Challenge 3: Type abilities of objects.} 
All Move objects can have several associated abilities to constrain how they can be used and persisted. For example, the struct type \verb|VeryAble| in Figure~\ref{fig:intro_example} has \verb|drop| and \verb|store| abilities while \verb|Receipt| has none of them.
The \verb|drop| and \verb|store| are two abilities that may affect the validity of a transaction.
The \textit{store} ability determines if an object can be stored in persistent storage, while the \textit{drop} ability allows an object to be dropped without being used.
An object without \verb|drop| and \verb|store| abilities, such as \verb|Receipt|, can not be transferred to other accounts and thus must be consumed by \emph{one and only one function call} within the same transaction.
Such objects are usually called \textbf{``Hot Potato''} and are typically used for implementing flash loans~\cite{flashloan}.
It enables users to perform permissionless and collateral-free loans and ensures that the loans are repaid in a single transaction, \issta{where a transaction on Move refers to a sequence of Move function calls as defined in Section~\ref{sec:preliminaries}}.
When generating a transaction, a fuzzer must ensure that all produced ``Hot Potato'' objects are properly handled.
For example, if a fuzzer creates a function call \verb|loan<T>()| in a transaction, then the created object of type \verb|Receipt<T>| should be passed as input to another function, in this case, only \verb|repay<T>| to consume it.
Such ``Hot Potato'' objects introduce constraints across different function calls.

\para{Our contributions.}
In this paper, we propose a \emph{\wanxu{dependency graph}}-guided Move fuzzing framework, \tool.
Given a Move smart contract, we build a \emph{\wanxu{dependency graph}} to model all the type constraints. The \wanxu{dependency graph} will later be used to guide the generation and mutation of transactions during fuzzing.
For example, Figure~\ref{fig:intro_typegraph} shows a simplified \wanxu{dependency graph} for the smart contract in Figure~\ref{fig:intro_example}. Nodes in the graph represent different object types or functions, while the annotated edges maintain different relations and constraints among them.
In order to further overcome tricky checks in Move smart contracts, we also design and implement a concolic executor in \tool.

We extensively evaluate \tool on a large-scale dataset provided by a \rev{well-known Web3 audit company} and a real-world dataset collected by an empirical work~\cite{movescan}.
The results are very promising: \tool is able to detect 100\% critical vulnerabilities and \zq{79\%} major vulnerabilities when cross-compared with manually audited results by human experts.
\zq{We also selected two notorious incidents in the Move ecosystem, \cetus with \$200 million loss and \nemo with \$2.6 million loss. \tool successfully reproduced both incidents with full exploits.}
We envision that this work bridges the gap between Move's static safety guarantees and dynamic testing.

\section{Motivation}\label{sec:motivation}
In this section, we motivate our design by first showing the deficiency of existing fuzzers on Move smart contracts, then empirically demonstrating the criticality of handling the special constraints in Move, and lastly exemplifying how \tool can generate valid transactions with the guidance of the \wanxu{dependency graph}.

\subsection{Deficiency of Existing Fuzzing Tools on Move}\label{sec:motivation-existing}

\begin{figure}[tp]
     \centering
    \begin{minipage}[b]{0.4\linewidth}
        \begin{subfigure}{\linewidth}
        \raggedright
            \begin{minted}[frame=single,framesep=1pt,bgcolor=mygray,xleftmargin=1em,fontsize=\footnotesize,escapeinside=@@]{rust}
v1, v2 = loan@\textbf{\color{red}\texttt{<T>}}@(100);
            \end{minted}
            \vspace{-5pt}
            \caption{Both \ityfuzz and \suifuzzer cannot handle type parameter \textbf{\color{red}\texttt{<T>}}.}
            \label{fig:motivation_both_fail}
        \end{subfigure}

        \vspace{5pt}
        \begin{subfigure}{\linewidth}
            \begin{minted}[frame=single,framesep=1pt,bgcolor=mygray,xleftmargin=1em,fontsize=\footnotesize,escapeinside=@@]{rust}
v1, @\textbf{\color{red}\texttt{v2}}@ = loan(100);
v3, v4 = loan(100);
repay(v1, v4);
            \end{minted}
            \vspace{-5pt}
            \caption{After removing type parameters, \suifuzzer still missed the ``Hot Potato'' \textbf{\color{red}\texttt{v2}}.}
            \label{fig:motivation_no_para}
        \end{subfigure}

        \vspace{5pt}
        \begin{subfigure}{\linewidth}
            \begin{minted}[frame=single,framesep=1pt,bgcolor=mygray,xleftmargin=1em,fontsize=\footnotesize,escapeinside=@@]{rust}
v1, v2 = loan<USDC>(100);
repay<USDC>(v1, v2);
            \end{minted}
            \vspace{-5pt}
            \caption{\tool successfully generates a valid transaction by correctly instantiating type parameters and handling ``Hot Potato''.}
            \label{fig:motivation_our}
        \end{subfigure}
        \caption{Fuzzer-generated transactions.}
    \end{minipage}
     \hfill
     \begin{minipage}[b]{0.48\linewidth}
         \centering
            \includegraphics[width=\linewidth]{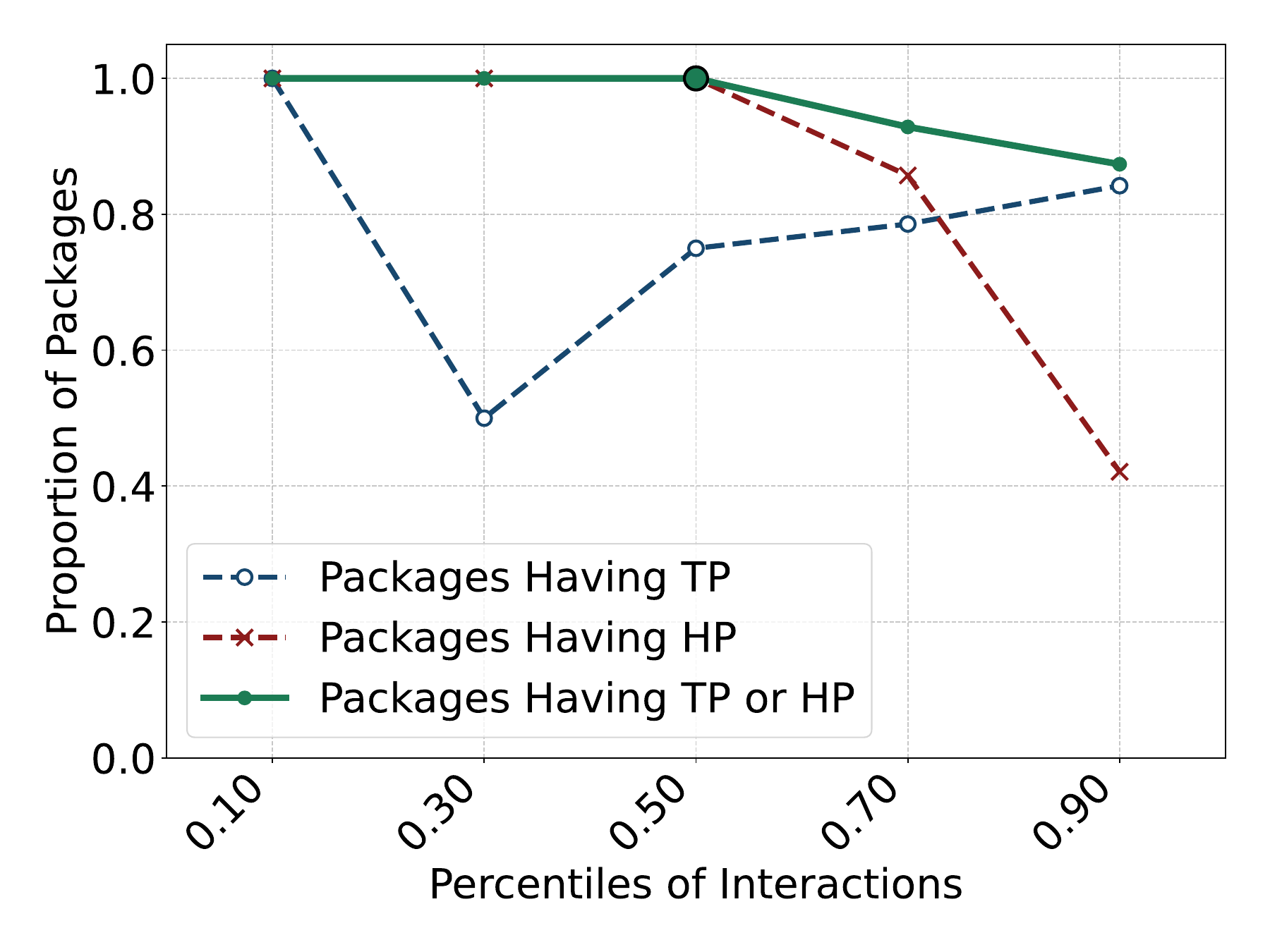}
            \caption{Ratio of the packages having type parameters (denoted as TP) and ``Hot Potato'' (denoted as HP)}
            \label{fig:popularity}
     \end{minipage}
\end{figure}

\ityfuzz and \suifuzzer are two state-of-the-art smart contract fuzzers that can be used on Move smart contracts.
To illustrate their deficiency, we create a Move smart contract with the example shown in Figure~\ref{fig:intro_example}. This example, although constructed, includes the main and commonly seen features in Move smart contracts.
We try to run both \ityfuzz and \suifuzzer on this smart contract. 
Unfortunately, all the transactions generated by them have type-mismatch errors and \rev{thus are rejected without execution}, such as the one shown in Figure~\ref{fig:motivation_both_fail}.
The root cause is that neither of them supports type parameters, which are used in the structs and functions in our example.

We further simplify the smart contract by removing all the type parameters and the type \verb|Coin<T>| from the standard library.
However, \suifuzzer still fails to generate any valid transactions, as shown in Figure~\ref{fig:motivation_no_para}. The reason is two-fold. First, it will internally forge objects by filling random bytes, which can easily violate the type safety rule of Move. Second, it is not aware of the type abilities of objects and thus ``Hot Potato'' objects are not properly handled.
\ityfuzz generates some valid transactions in this simplified smart contract by randomly mutating function calls until all ``Hot Potato'' objects are handled. This random mechanism is highly inefficient and can only produce ``Hot Potato'' objects without type parameters.
Although existing fuzzers fail to generate valid transactions on smart contracts with type parameters and ``Hot Potato'', we still lack an understanding of how common such features are in the real world.
In order to measure the prevalence of type parameters and ``Hot Potato'', we collected all \emph{two billion} transactions on the Sui blockchain from November 2024 to November 2025. 
Since Move smart contracts are organized as packages, we obtain the source packages that each transaction uses. 
For all the packages, we count the number of packages with type parameters and ``Hot Potato''.
Figure~\ref{fig:popularity} shows the percentage of packages that have type parameters, ``Hot Potato'', and at least one of them.
Packages are ranked based on the number of \zq{Move function calls that use them}.
The result shows that among the top 90\% most-used packages, more than 80\% of them have at least one type parameter or ``Hot Potato'', indicating the prevalence of such Move features in the real-world deployment.


\definecolor{mygray}{gray}{0.9}

\subsection{Effectiveness of \tool}

\tool overcomes the limitations of previous fuzzers by implementing type-aware transaction generation and mutation.
With the statically built \wanxu{dependency graph} for a smart contract, \tool can always generate valid transactions.
For example, to generate a transaction for Figure~\ref{fig:intro_example}, \tool first queries the \wanxu{dependency graph} to understand that the output object of \verb|loan<T>()| contains a ``Hot Potato'' that can be consumed by \verb|repay<K>()|. Since there are no additional type constraints, the two type parameters \verb|T| and \verb|K| can be instantiated with the same concrete type \verb|USDC|. 
With all the information obtained from the \wanxu{dependency graph}, \tool can always produce valid transactions for testing this smart contract, as shown in Figure~\ref{fig:motivation_our}.

\section{Methodology}\label{sec:method}

In this section, we first introduce some preliminary definitions to facilitate our methodology description, and then we introduce the high-level workflow of \tool. We elaborate on the technical details of the core components in the subsequent parts.

\subsection{Preliminaries}\label{sec:preliminaries}
We make the following definitions to facilitate our description of the methodology.

\begin{definition}[Object]
\shaohua{An object refers to a typed entity. Each object is an instance of a Move struct or enum type that possesses certain abilities.}
\end{definition}
Objects can be created by a transaction or persisted in on-chain blockchain storage. The \verb|v1| and \verb|v2| in Figure~\ref{fig:motivation_our} are two objects created by the transaction. On-chain objects are created and stored by some transactions. A new transaction can use on-chain objects \zq{owned by the transaction sender} or shared objects.
On-chain objects are non-``Hot Potato'' by definition, and thus a fuzzer can directly use them as long as the type is matched.

\begin{definition}[Transaction]
A transaction $S$ in Move is a sequence of function calls without control flow, \issta{where each function call may use primitive values or existing objects as parameters}.
\end{definition}
On both Sui and Aptos, the two prevailing blockchains that adopt the Move language, a transaction can carry multiple Move function calls with slightly different implementation details.
Sui supports \textit{Programmable Transaction Block} (PTBs) composed of a sequence of function calls, each of which can use either an existing object or a value returned from previous Move function calls.
Aptos allows users to ad-hoc compile and run a piece of Move code as a transaction, named \textit{Script}, instead of deploying a Move contract, which is essentially a superset of \textit{Programmable Transaction Block} with arbitrary control flows.
Both mechanisms enable a transaction to do multiple Move function calls in an atomic fashion.
Thus, without loss of generality, a transaction in this paper is defined as a sequence of function calls without control flows.

\begin{definition}[Object consumption]
For a function call $f$, if object $o$ is part of its input, we say the function $f$ consumes object $o$.
\end{definition}

\begin{definition}[Object production]
For a function call $f$, if object $o$ is part of its output, we say the function $f$ produces object $o$.
\end{definition}

These two definitions are important for us to describe how we handle ``Hot Potato'' objects during transaction generation and mutation.


\subsection{\tool in a Nutshell}\label{sec:nutshell}

\tool is essentially a gray-box fuzzing method. It shares the typical components of the conventional fuzzers, such as seed initialization and seed mutation. We briefly discuss the fuzzing loop and postpone the technical details to the next sections.

\begin{figure*}[t]
\begin{subfigure}{0.6\textwidth}
    \begin{minted}[xleftmargin=1em,numbersep=3pt,fontsize=\footnotesize,linenos,escapeinside=@@]{rust}
module pool;
public struct Receipt<T> {...}

public fun loan<T>(amount: u64): (Coin<T>, Receipt<T>) {...}

public fun repay<T>(coin: Coin<T>, receipt: Receipt<T>) {
    // Function to repay flash loan with coins.
    let Receipt {
        amount,
        fee
    } = receipt;  // unpack the receipt and consume it.
    let paid = ...// extract from the input coin
    assert!(paid == amount + fee, "not enough repay");@\label{line:fee}@
    ... // potentially vulnerable path
}
public fun swap<T1, T2>(coin: Coin<T1>): (Coin<T2>) {...}

public fun split_coin<T>(coin: Coin<T>, split: u64)
    : (Coin<T>, Coin<T>) {...}
    \end{minted}
    \caption{A constructed Move smart contract.}
    \label{fig:approach_example}
\end{subfigure}
\hfill
\begin{subfigure}{0.35\textwidth}
    \centering
    \includegraphics[width=\linewidth]{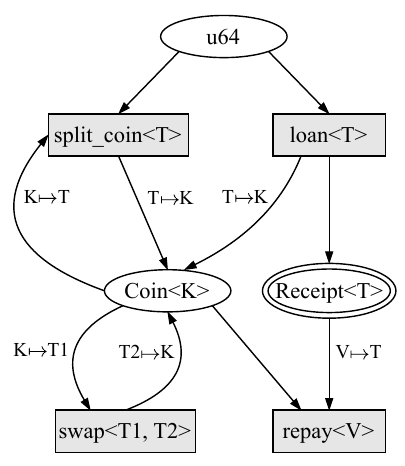}
    \caption{Our constructed \wanxu{dependency graph} for the code in (a).}
    \label{fig:approach_typegraph}
\end{subfigure}
    \caption{A constructed Move smart contract (left) and the built \wanxu{dependency graph} for it (right).}
\end{figure*}

\newcommand{\objpool}{$\texttt{Pool}_{obj}$\xspace}
\begin{itemize}[leftmargin=20pt]
    \item \textbf{Step 1: Seed initialization.} 
    Initially, the seed corpus contains no transactions. Since smart contracts are stateful programs, transactions may use on-chain objects. We thus initialize an object pool, denoted as \objpool, by crawling all on-chain objects that are related to the target smart contract. \objpool will be initialized when the fuzzer starts and be reset to the initialized state before generating a new transaction.

    \item \textbf{Step 2: \wanxu{Dependency graph} construction.} Since Move is a strongly typed language, we need to ensure the type validity of all the generated/mutated transactions. To this end, we construct a \wanxu{dependency graph} $G$ to capture all necessary type-related information, which can then be used to guide the generation and mutation of transactions. Our usage of the \wanxu{dependency graph} $G$ ensures that all the transactions produced are well-typed. (Details in Section~\ref{sec:type})

    \item \textbf{Step 3: Transaction generation and mutation.} In each iteration, \tool either randomly generates a new transaction or mutates an existing seed transaction. 
    \begin{itemize}[leftmargin=10pt]
        \item [$\circ$] \emph{For random generation}, \tool first selects a random function node in the \wanxu{dependency graph}. Then, \tool walks on the \wanxu{dependency graph} to find additional function nodes that are required to make a valid transaction. For each function call, \tool guarantees that each input object is produced by some earlier function or from \objpool, and each output ``Hot Potato'' object is properly consumed. (Details in Section~\ref{sec:fuzzing})

        \item [$\circ$] \emph{For random mutation}, \tool first selects a seed transaction from the seed pool and then mutates it by modifying constant values or adding/removing function calls. Since newly added or removed function calls may break type validity, \tool again uses the \wanxu{dependency graph} to maintain validity. (Details in Section~\ref{sec:fuzzing})

        \item [$\circ$] \emph{Concolic execution.} The above generation and mutation processes only care about the type validity of a transaction. For primitive types such as \verb|u64| that are used as function inputs, they all choose or mutate their values randomly. However, such a random strategy is hard to cover hard-coded conditions or checks that are prevalent in smart contracts. For example, line~\ref{line:fee} in Figure~\ref{fig:approach_example} can only be triggered when the inputs satisfy the value constraint, which is nearly impossible to reach with random values.
        To tackle this challenge, we design and implement a concolic executor based on the Move Virtual Machine. 
        (Details in Section~\ref{sec:concolic})
    \end{itemize}

    \item \textbf{Step 4: Transaction execution and feedback collection.} The generated new transaction will then be executed on the Move Virtual Machine (MoveVM). If the transaction increases the code coverage\footnote{We use the standard branch coverage in our implementation.}, it will be added to the seed corpus for further mutation. 
    If the execution reveals an oracle violation, it will be reported as a potential vulnerability. Different from C/C++ software, which typically has a standard oracle, such as buffer overflow, there is no standard oracle in smart contracts, and the oracle problem remains an unsolved problem~\cite{defiranger,verite,ityfuzz}.
    In our implementation, we provide six pre-defined common oracles that we observe in Move smart contracts. Additionally, we expose a user-defined oracle interface that enables the tool users to define their own oracles. (Details in Section~\ref{sec:oracle})

\end{itemize}

\subsection{\wanxu{Dependency Graph} Construction}\label{sec:type}
Given a target smart contract, our first step is to construct its \wanxu{dependency graph} $G = (V, E)$, where $V$ is the set of nodes representing either a type (struct or primitive type) or a function definition, and $E$ is the set of labeled edges. 
Figure~\ref{fig:approach_example} shows an example Move smart contract. It contains three types, \ie, primitive type \verb|u64|, newly created object type \verb|Receipt<T>|, and object type \verb|Coin<T>| from the standard library, and four function definitions.
Figure~\ref{fig:approach_typegraph} shows the \wanxu{dependency graph} we constructed for this smart contract.
There are three different kinds of nodes in a \wanxu{dependency graph}:
\begin{itemize}[leftmargin=20pt]
    \item \textbf{Default type node}: A type node that is a primitive type, such as \verb|u64|, or a struct type with at least \verb|drop| or \verb|store| capability, such as \verb|Coin<T>|.
    \item \textbf{Hot potato type node}: A type node where the type has neither \verb|drop| nor \verb|store| ability, such as \verb|Receipt<T>|.
    \item \textbf{Function node}: A function node that represents a function definition, such as \verb|loan<T>()|.
\end{itemize}

Move smart contracts usually rely on standard libraries or third-party smart contracts. Consequently, a smart contract may use types defined somewhere else. In order to build a \wanxu{dependency graph} with complete information, we need to traverse all the libraries and smart contracts that are referenced by our target smart contract. {In our implementation, \tool supports building the \wanxu{dependency graph} for a group of smart contracts at the same time.}
\wanxu{Transaction generation can therefore compose type-compatible function-call sequences across module and package boundaries, as long as the corresponding functions and types are present in the graph and the required objects are available in \objpool.}
\wanxu{This is critical for real-world DeFi contracts, where one package often consumes objects or price/oracle outputs produced by another package.}

Our \wanxu{dependency graph} is inspired by the \emph{signature graph} introduced in program synthesis~\cite{signaturegraph} and the \emph{API graph} used in API-driven program synthesis~\cite{apigraph}.
\wanxu{It encodes the input and output type relations of functions, type-substitution constraints, and ability constraints needed for executable Move transactions.}
Our \wanxu{dependency graph} construction initializes the \wanxu{dependency graph} $G$ with every used primitive type as a default type node $n_t$.
Then, it analyzes all the struct types.
If the struct has at least \verb|drop| or \verb|store| ability, it is added as a default type node $n_t$, illustrated as ovals in Figure~\ref{fig:approach_typegraph}; Otherwise, it is added as a ``Hot Potato'' type node $n_h$, illustrated as double-circled ovals in Figure~\ref{fig:approach_typegraph}.
Then, for each function definition $f$ in the target smart contract, we proceed as follows.
\begin{itemize}[leftmargin=20pt]

    \item Add $f$ as a function node in the \wanxu{dependency graph} $G$.
    
    \item For each input type $t_i$,
        \begin{itemize}[leftmargin=20pt]
        
            \item[$\circ$] If $t_i$ is a non-polymorphic type, add the edge ${t_i} \rightarrow f$ that connects the type node ${t_i}$ and the function node $f$ to the \wanxu{dependency graph} $G$. For example, \verb|u64|$\rightarrow$\verb|loan<T>()| is added as \verb|u64| is the input type of \verb|loan<T>()|.
            
            \item[$\circ$] If $t$ is a polymorphic type with type parameter \verb|K|, \ie, $t\verb|<K>|$, add the edge $t\verb|<K>| \xrightarrow{K \mapsto T} f$ to the \wanxu{dependency graph} $G$, where the edge connects the type node $t\verb|<T>|$ and the function node $f$. The annotation $\texttt{K} \mapsto \texttt{T}$ means type substitution.
            Since there is no subtyping\footnote{Although \texttt{tuple} in Move supports a very limited subtyping rule, it cannot be used as a type parameter~\cite{movesubtyping}.} in Move, type substitution can be greatly simplified. We use $\texttt{K} \mapsto \texttt{T}$ to represent that \verb|T| will be instantiated with the concrete type that instantiates \verb|K|.
            For example, \verb|Coin<K>|$\xrightarrow{\texttt{K} \mapsto \texttt{T}}$\verb|split_coin<T>| is added according to this rule. During transaction generation, when we instantiate \verb|Coin<K>| as \verb|Coin<u64>| and pass it as input to \verb|split_coin<T>()|, the function \verb|split_coin<T>()| also instantiates its type parameter \verb|T| to \verb|u64|.
        \end{itemize}
    
    \item For each output type $t_o$, similar to the input type, we add the edge and the annotation depending on whether it is a non-polymorphic or a polymorphic type.
\end{itemize}

This complete \wanxu{dependency graph} will enable us to know which type can be produced or consumed by which functions, thus facilitating our later transaction generation and mutation.

\begin{figure*}[t]
    \centering
    \includegraphics[width=1.0\linewidth]{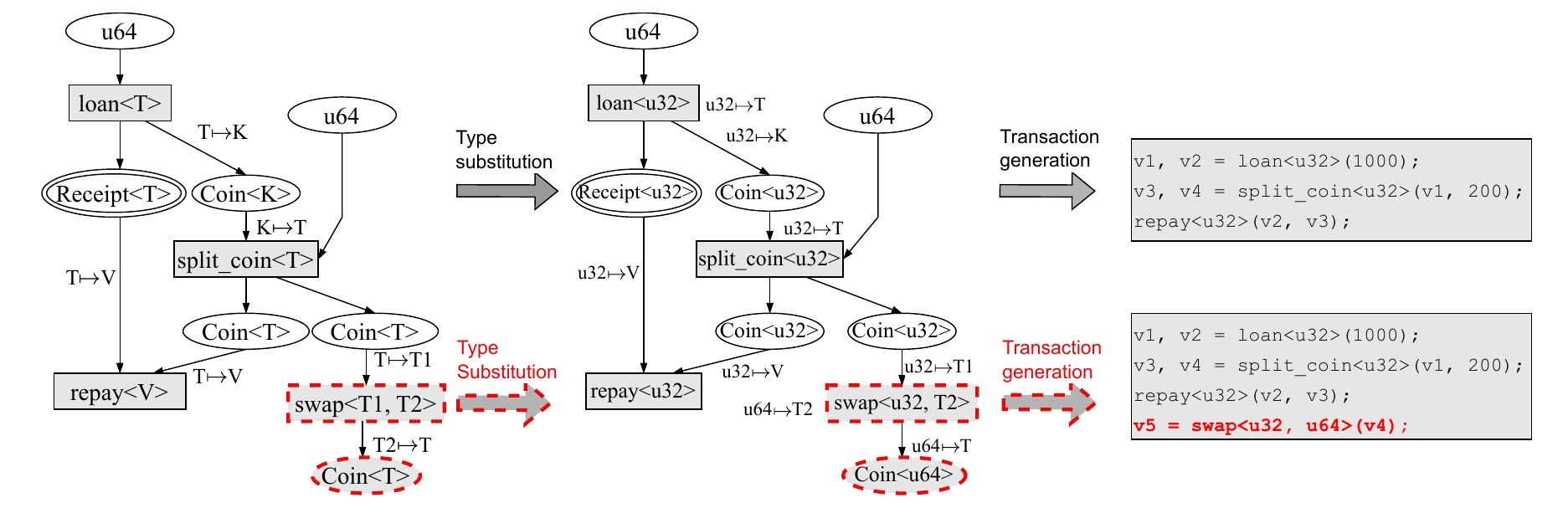}
    \caption{Illustrative transaction generation starting from a graph trace (left), then concretizing type parameters by type substitution (middle), and finally instantiating a transaction (right). The {\color{red} dashed nodes} demonstrate the mutation process of graph trace extension.}
    \label{fig:programgeneration}
\end{figure*}

\subsection{Transaction Generation and Mutation}\label{sec:fuzzing}

\newcommand{\typegraph}{$G$\xspace}

\newcommand{\gtrace}{$G_{trace}$\xspace}

There are two ways of producing a transaction in \tool, namely generation and mutation. They share the same high-level idea but differ in the implementations. Below we introduce the details.

\para{Transaction Generation.}
The high-level idea of generating a transaction is to search for a graph trace \gtrace in the \wanxu{dependency graph} \typegraph such that we can generate a valid transaction by instantiating \gtrace. It has three main steps.

\begin{itemize}[leftmargin=15pt]
    \item \textbf{Step 1: Graph trace construction.} The purpose is to walk on the \wanxu{dependency graph} to find a graph trace \gtrace, whose \issta{source nodes (no incoming edges) and sink nodes (no outgoing edges)} are all \emph{default type nodes}. Theoretically, we could start from any node in \typegraph to find such a trace. But to make our algorithm run faster, we start from a function definition that only accepts void or primitive types as input. 
    This choice does not affect the overall outputs of the algorithm.
    The reason is that, unless using shared objects from \objpool, the first function call in a valid transaction must create objects with void or primitive types.
    For example, the figure on the left in Figure~\ref{fig:programgeneration} shows one graph trace obtained by walking on the example \wanxu{dependency graph} in Figure~\ref{fig:approach_typegraph} starting from \verb|loan<T>|.

    \item \textbf{Step 2: Type substitution.} Since Move accepts type parameters, before generating the final transaction, we need to find the concrete types for each type parameter in the graph trace. Since types in a Move smart contract are usually limited (fewer than 10), we use a heuristic-based method to iteratively find suitable types for each type parameter.
    For example, the middle figure in Figure~\ref{fig:programgeneration} shows the graph traces after type substitution. 

    \item \textbf{Step 3: Transaction generation.} With a concretized graph trace, we can then generate the final transaction by instantiating it. The process starts from all type nodes without incoming edges in the graph trace and adopts a breadth-first enumeration to generate each function call. 
    For example, in Figure~\ref{fig:programgeneration}, we start from the type node \verb|u64| and generate the function call \verb|loan<u32>(1000)|, where \verb|1000| is a random value generated according to its type. Then we generate the \verb|split_coin<u32>()| and \verb|repay<u32>()| function calls to consume all objects created by \verb|loan<u32>()|.
\end{itemize}

The above process can be repeated multiple times, creating a new graph trace each time.
Even for the same starting node, the algorithm can generate multiple graph traces, while each trace can be instantiated differently when using different concrete types to substitute type parameters.
Note that for primitive types, such as \verb|u64|, we use random values for them when generating a transaction, such as \verb|1000| for \verb|loan<T>()| in Figure~\ref{fig:programgeneration}. Although using random values is a common practice in fuzzing, it can be very hard for them to reach certain code path that requires strict condition checks. To overcome this limitation, we design a concolic executor in \tool. We will introduce more details in Section~\ref{sec:concolic}.

\para{Transaction Mutation.}
Transactions that increase code coverage will be added to the seed corpus for further mutation.
When a transaction $S$ is selected for mutation, we first recover its graph trace \gtrace and use one or several of the following mutators to produce a new transaction.

\begin{itemize}[leftmargin=15pt]
    \item \textbf{Mutator 1: Value mutation.} For primitive values in $S$, we mutate them by doing \rev{AFL-style havoc mutation\cite{libafl}, \ie, randomly mutating several bits or bytes in $S$.}

    \item \textbf{Mutator 2: Graph trace extension.} Randomly select a type node in \gtrace that is not consumed and extend the graph trace by walking on the \wanxu{dependency graph} $G$ until we get a new valid graph trace $G_{trace}'$, then instantiate $G_{trace}'$ to a transaction.
    For example, the dashed nodes in Figure~\ref{fig:programgeneration} illustrate this mutation. We first select the type node \verb|Coin<T>| that is not consumed yet in the original \gtrace. Then we walk on the \wanxu{dependency graph} \typegraph to add the function node \verb|swap<T1,T2>()| to the \gtrace. We will then do type substitution on it and finally instantiate a mutated transaction from it.
    Note that it is possible that there is no such type node in \gtrace. In this case, this mutator will be skipped.

    \item \textbf{Mutator 3: Function call insertion.} Randomly select a new function node $f'$ in the \wanxu{dependency graph} and construct a graph trace $G_{trace}'$ by trying to connect $f'$ to \gtrace. We select a type node in \gtrace that is not consumed as the target, and then try to search for the connection path by random walk. If this search process succeeds, this path will be added to \gtrace, and then we will include additional necessary nodes to make this new graph trace complete.
    
    If this search process fails, meaning that there is no valid type node in the original \gtrace or there is no such path, we invoke the \emph{transaction generation} algorithm with $f'$ as the starting node to generate a transaction. This transaction will be appended to the original transaction to get a new transaction.
    For example, the dashed nodes in Figure~\ref{fig:programgeneration} can also illustrate this process if we view \verb|swap<T1,T2>| as the selected function node.

    \item \textbf{Mutator 4: Function call removal.} Randomly select a function node in \gtrace and then remove it from \gtrace. This removal may break the validity of \gtrace, and thus we keep removing necessary nodes until the remaining nodes constitute a valid graph trace $G_{trace}'$. Finally, we instantiate $G_{trace}'$ into a new transaction. 
\end{itemize}

For a seed transaction $S$ and its graph trace \gtrace, more than one of the mutators may be applied. If this resulting transaction $S'$ triggers new code coverage, $S'$ will be added into the seed corpus for further mutation.

\subsection{Concolic Execution}\label{sec:concolic}
The concolic executor is an important component for ensuring \tool can reach critical paths in the target Move smart contracts.
\wanxu{We implement the concolic executor on top of the MoveVM, which provides the runtime values of variables and stacks.
In particular, stack values in MoveVM are also strongly typed, making it relatively easy for us to model symbolic expressions, avoiding much of the object recovery and byte-level memory reasoning required for C/C++ symbolic execution~\cite{symhard}.}
\wanxu{We invoke concolic execution after every concrete transaction execution, with a 500~ms timeout per transaction.}
\wanxu{This bound prevents solver calls from dominating the fuzzing loop while still giving the solver enough budget to repair common arithmetic, cast, and vector-access constraints.}
\wanxu{If the solver times out or returns unsatisfiable, \tool applies ordinary value mutation.}

\para{\wanxu{Constraint collection.}}
\wanxu{For a concrete transaction, \tool first traces its MoveVM bytecode execution and records the path constraints that guard the observed execution.}
\wanxu{During execution, the symbolic expressions are propagated through local variables, stack values, and function-call boundaries so that constraints can relate inputs used by different calls in the same transaction.}

\wanxu{The collected constraints mainly cover four cases.}
\begin{itemize}[leftmargin=20pt]
    \item \textbf{\wanxu{Branch and assertion predicates.}} \wanxu{\tool records the predicates that determine whether an execution enters a guarded path, including Move assertions and conditional branches.}
    \item \textbf{\wanxu{Arithmetic operations.}} \wanxu{\tool records arithmetic constraints for operations such as addition and multiplication. Since Move arithmetic aborts on overflow, \tool also adds range constraints so that solver-produced values keep the mutated transaction executable.}
    \item \textbf{\wanxu{Cast operations.}} \wanxu{MoveVM uses explicit cast bytecodes, such as \texttt{castu16}, to change integer widths. If the runtime value exceeds the destination type range, execution aborts; therefore, \tool encodes source-to-destination range constraints for casts.}
    \item \textbf{\wanxu{Vector access.}} \wanxu{When a transaction indexes a vector, MoveVM aborts if the index is out of bounds, so \tool records index constraints for vector accesses.}
\end{itemize}

\para{\wanxu{Concolic mutation.}}
\wanxu{After collecting a transaction's constraints, \tool randomly selects one or more branch predicates and flips their conditions while preserving other constraints.}
\wanxu{If the solver finds satisfying assignments, \tool replaces the corresponding concrete primitive inputs in the original transaction and obtains a new transaction for execution.}


\section{Implementation}\label{sec:impl}\label{sec:oracle}

We build \tool from scratch, totaling 17k lines of Rust code.
Specifically, we spend 4.1k lines to support forking a specific on-chain checkpoint to enable fast and local on-chain data access.
With this support, \tool can initialize the object pool with on-chain objects. 
The \wanxu{dependency graph} building, graph tracing, and transaction generation and mutation take 3.7k lines, while the concolic execution takes 2k lines.

\para{Oracle.}
To detect bugs in Move smart contracts, we require oracles to determine whether each transaction is executed correctly.
Unfortunately, designing an automated, general-purpose oracle for fuzzing smart contracts remains a long-standing open problem~\cite{defiranger}. The primary reason is that oracles often depend on the target smart contract's business model or semantics. One can think of it as detecting functional bugs in general software, which has remained unsolved for decades.
Therefore, following a similar practice in previous smart contract fuzzers like Foundry\cite{foundry} and Echidna\cite{echidna}, \tool provides complementary Move libraries to enable users to implement custom oracles in pure Move language by emitting specific events when invariants are violated. We will introduce more details in the evaluation in Section~\ref{sec:eval}.

A recent empirical study~\cite{movescan} defines some common oracles for Move smart contracts. These oracles are mostly for checking mild issues in a smart contract, such as \emph{unnecessary type conversion}, which may waste some gas.
We implement and refine all five oracles from \cite{movescan} that can be used during fuzzing in \tool. Additionally, we implement a new general oracle ``\emph{earning profits}'' for detecting arbitrary earnings. Below is a list of the details about these oracles.
\wanxu{Although we implement these predefined oracles, oracle design is orthogonal to the core contribution of this paper.}
\wanxu{They are the default checks when users do not provide protocol-specific invariants, and they mainly demonstrate that the transaction-generation engine can drive both generic and custom bug detectors.}

\begin{itemize}[leftmargin=20pt]
    \item \emph{Infinite Loop:} detect infinite loops in a contract. \tool checks if the expression and value of a branch condition remain unchanged until the execution is out of gas.
    \item \emph{Precision Loss:} check both operands of an arithmetic operation to determine if precision loss happens, \eg, \verb|5/2 => 2| instead of \verb|2.5|. \issta{\tool utilizes the concolic execution to precisely track the symbolic expressions rather than pattern matching~\cite{movescan}.}
    \item \emph{Unnecessary Type Conversion:} check if the type of the stack top during execution is the same as the cast destination type.
    \item \emph{Unnecessary Bool Judgment:} check if at least one of the operands of \textit{EQ} or \textit{NEQ} is a constant boolean value. In this case, the comparison is not necessary because the other operand itself could be used to determine the condition by either \textit{BrFalse} or \textit{BrTrue}.
    \item \emph{SHL Overflow:} detect left shift overflow. \issta{\tool tracks the runtime values instead of static analysis to ensure no false positives.}
    \item \emph{Earning Profits:} check if the amount of coins owned by the transaction sender increased after the execution. Since we do not mint additional coins in generated transactions, the amount of coins should never be increased. Otherwise, we can earn profits without any cost, indicating serious issues.
\end{itemize}

\para{Supporting different blockchains.} To date, the codebases and semantics of Sui and Aptos, two prevailing blockchains using Move \issta{that once shared code}, have diverged significantly. This leads to distinct features with entirely different interfaces.
For different blockchains, most changes are in coverage instrumentation and instruction tracing, totaling 483 and 638 lines of difference in the two blockchains, respectively.

\section{Evaluation}\label{sec:eval}

\newcommand*{\dsb}{\textbf{DS2}\xspace}
\newcommand*{\dsa}{\textbf{DS1}\xspace}

\newcommand*{\rqone}{\textbf{RQ1}\xspace}
\newcommand*{\rqtwo}{\textbf{RQ2}\xspace}
\newcommand*{\rqthree}{\textbf{RQ3}\xspace}
\newcommand*{\rqfour}{\textbf{RQ4}\xspace}

\newcommand*{\noconcolic}{\textbf{\tool-NCE}\xspace}
\newcommand*{\notg}{\textbf{\tool-NTG}\xspace}
\newcommand*{\noty}{\textbf{\tool-NTY}\xspace}
\newcommand*{\noall}{\textbf{\tool-DUMB}\xspace}

In this section, we evaluate \tool on datasets collected from real-world auditing and compare it with other related fuzzers. We aim to answer the following research questions in this section.

\begin{itemize}[labelwidth=!, labelindent=5pt, itemsep=3pt, topsep=2pt, leftmargin=15pt]
    \item \textbf{RQ1 (Bug Reachability).} \emph{Is \tool effective at \issta{reaching vulnerabilities} in real-world Move smart contracts?} 
    \item \textbf{RQ2 (Coverage and throughput).} \emph{Can \tool achieve higher code coverage and throughput?}
    \item \textbf{RQ3 (Ablation study).} \emph{How do the key components in \tool affect the overall effectiveness?} 
    \item \textbf{RQ4 (Incident Reproduction).} \issta{\emph{Can \tool reproduce previous on-chain incidents?}}
    \item \textbf{RQ5 (Real-world Application)}. \issta{\emph{Can \tool find new bugs in real-world smart contracts?}}
\end{itemize}

\smallskip
\noindent\textbf{Dataset Construction.} 
Two datasets are used in our evaluation.
The first dataset, \dsa, is assembled from audit reports provided by \rev{a well-known Web3 auditing company}.
In this dataset, all Move smart contracts are manually audited by human experts, with detailed vulnerability information and further confirmed by project developers.
\wanxu{These reports provide ground-truth vulnerability locations and violated invariants, which serve as the oracle used in \dsa. Due to the availability of the oracle, \dsa is used to evaluate oracle-guided bug reachability rather than oracle synthesis.} Although it is possible that some vulnerabilities escape human experts' audit, it does not affect the evaluation results of our experiments.
Each vulnerability is labeled based on its severity \issta{acknowledged by the developers}.
\issta{As the audit reports define}, \textit{Critical} vulnerabilities mostly lead to fund loss.
\textit{Major} vulnerabilities usually relate to the design flaws of the business models or improper access control that are not directly exploitable.
\textit{Medium} severity usually indicates a potentially incorrect setup or configuration causing minor losses or unexpected results, \ie, non-security issues.
We exclude the \textit{Minor} and \textit{Informational} vulnerabilities because they are either code style issues or harmless defects, such as gas waste, and the audit reports mark these issues as not compulsory to fix. 
We select projects that contain at least one \textit{Medium} vulnerability in the past year, which yields 37 projects containing \rev{960} smart contracts with 9 \textit{Critical}, 53 \textit{Major}, and 47 \textit{Medium} vulnerabilities.\footnote{\rev{Following \movescan convention, we deem a Move module as a smart contract and one Move project usually contains multiple Move modules.}}
The second dataset, \dsb, is from \movescan~\cite{movescan}.
It contains 72 open-source Move \rev{projects}, totaling \rev{387} smart contracts on both Sui and Aptos.\footnote{We exclude 1 incorrectly labeled project, 1 project that deletes its repository, and 3 projects that no longer build.}.

\issta{As we discussed earlier in Section~\ref{sec:oracle}, designing an automated and general-purpose oracle for fuzzing smart contracts remains a long-standing open problem~\cite{defiranger}. Luckily, we have the ground-truth audit reports for \dsa. We thus implement the oracle for each smart contract, averaging 15 lines per vulnerability.
To avoid overfitting, we invite two additional Move auditors with at least 1 year of full-time experience to review the implementation. \dsb does not have ground truth about vulnerabilities, and we thus rely on our predefined oracles.}
\wanxu{The auditors use each audit finding to derive the violated invariant and the relevant public entry functions or state variables, without directly using audit PoCs as oracle code. Each custom oracle is a thin Move wrapper or invariant module that forwards the same arguments to the original function, records only the values needed by the audited invariant, and reports a violation through \tool's oracle event interface.}
\wanxu{We count a \dsa bug as found only when a generated transaction triggers the corresponding custom oracle violation.}
\wanxu{We map a finding to a ground-truth item by matching the target contract, the oracle wrapper, and the invariant described in the audit report; multiple transactions that trigger the same invariant are deduplicated as one found bug.}


\smallskip
\noindent
\textbf{Baseline Fuzzers.} We initially hoped to directly adopt two most relevant and state-of-the-art Move smart contract fuzzers, \ityfuzz\cite{ityfuzz} and \suifuzzer\cite{suifuzzer}, as the baselines.
Unfortunately, at the time of writing this paragraph, there have been 6 months and 19 months since the last maintenance of \ityfuzz and \suifuzzer, respectively.
Our preliminary efforts showed that they failed to start their fuzzing process in most smart contracts.
To evaluate them, we must first fix these two fuzzers.
Specifically, we add 503 lines to \ityfuzz to upgrade its Sui dependency, resolve various dirty hacks that cause MoveVM invariant violations, and fix a few bugs that cause segmentation faults and deadlocks.
For \suifuzzer, we also upgrade its Sui dependency, fix its build process, and extend its support for the \verb|init| function executed during Move contract deployment by an additional 381 lines of code.

\issta{
For \dsa, where we use custom oracles, we evaluate all baseline fuzzers and analyzers with our constructed oracles on it.
For \dsb, where we can only use predefined oracles, since only \movescan and \tool support such oracles, we evaluate them on it.}

\smallskip
\noindent
\textbf{Experiment Setup.} We conduct all the experiments on two machines, both equipped with an EPYC 7B13 processor and 1 TiB memory and running Ubuntu Server 24.04. We also bind each fuzzing campaign to a specific logical core and use a \textit{tmpfs} memory partition for any fuzzer output to avoid interference.
We repeat each fuzzing campaign five times, each for 12 hours. \zq{}

\subsection{RQ1: Bug Reachability}\label{sec:eval-end2end}

\begin{table*}[t]
\begingroup
\scriptsize

\centering
\setlength{\abovecaptionskip}{3pt}
\setlength{\belowcaptionskip}{0pt}
\caption{Overall bug-reachability results on \dsa. Per-project comparison between \tool and \ityfuzz. \rev{"MA", "ME" and "FP" are abbreviated for "Major", "Medium" and "False Positive", respectively.} \issta{Note \suifuzzer is evaluated but excluded because it does not yield any true positive findings.}}
\label{tab:rq1-bugs}
\setlength{\tabcolsep}{5pt}%

\renewcommand{\arraystretch}{1.05}
\begin{minipage}[t]{\textwidth}
\centering
\rowcolors{4}{lightgray!50}{}
\begin{tabular}{lccccc|lccccc}
\toprule
\multirow{2}{*}{\textbf{Target}} 
  & \multicolumn{4}{c}{\textbf{\tool}} 
  & \multirow{2}{*}{\textbf{\ityfuzz}} 
  & \multirow{2}{*}{\textbf{Target}} 
  & \multicolumn{4}{c}{\textbf{\tool}} 
  & \multirow{2}{*}{\textbf{\ityfuzz}} \\
  \cmidrule{2-5} \cmidrule{8-11}
 & \textbf{Total} & Critical & Major & Medium &
 &  & \textbf{Total} & Critical & Major & Medium & \\
\midrule
\shortstack[l]{8bc50340} & \textbf{5/5} & - & 2/2 & 3/3 & 0 &
\shortstack[l]{8f0c3e4c} & \textbf{5/6} & - & 0/1 & 5/5 & 0 \\
\shortstack[l]{c955a990} & \textbf{4/4} & 2/2 & 2/2 & - & 0 &
\shortstack[l]{d46b85fc} & \textbf{1/2} & 1/1 & - & 0/1 & 0 \\
\shortstack[l]{902aaf72} & \textbf{2/2} & - & 1/1 & 1/1 & 0 &
\shortstack[l]{c40e8d17} & \textbf{0/1} & - & - & 0/1 & 0 \\
\shortstack[l]{1e6f57ad} & \textbf{4/4} & - & 1/1 & 3/3 & 0 &
\shortstack[l]{4a93535d} & \textbf{1/2} & - & - & 1/2 & 1 ME (FP) \\
\shortstack[l]{916a5d9c} & \textbf{3/4} & - & 0/1 & 3/3 & 0 &
\shortstack[l]{f1b4dc69} & \textbf{2/3} & - & 2/3 & - & 0 \\
\shortstack[l]{73c07115} & \textbf{0/2} & - & 0/2 & - & 0 &
\shortstack[l]{1b894d8a} & \textbf{2/2} & - & - & 2/2 & 0 \\
\shortstack[l]{c52c5116} & \textbf{1/2} & - & 0/1 & 1/1 & 0 &
\shortstack[l]{60cb5d6d} & \textbf{3/5} & 1/1 & 2/3 & 0/1 & 0 \\
\shortstack[l]{d0c0ef13} & \textbf{4/4} & - & 1/1 & 3/3 & 0 &
\shortstack[l]{b3e614c7} & \textbf{2/2} & - & 1/1 & 1/1 & 0 \\
\shortstack[l]{e3564d47} & \textbf{1/1} & - & 1/1 & - & 0 &
\shortstack[l]{7d85e770} & \textbf{2/3} & - & 1/2 & 1/1 & 0 \\
\shortstack[l]{e8a96a03} & \textbf{2/2} & - & 1/1 & 1/1 & 0 &
\shortstack[l]{c2ade084} & \textbf{2/2} & - & 2/2 & - & 0 \\
\shortstack[l]{d834970c} & \textbf{2/2} & 1/1 & 1/1 & - & 0 &
\shortstack[l]{b830405a} & \textbf{3/4} & - & 1/2 & 2/2 & 0 \\
\shortstack[l]{074ef533} & \textbf{6/6} & 2/2 & 2/2 & 2/2 & 0 &
\shortstack[l]{c74a124c} & \textbf{4/4} & - & 3/3 & 1/1 & 0 \\
\shortstack[l]{52df0499} & \textbf{4/4} & - & 2/2 & 2/2 & 1 MA &
\shortstack[l]{d492c273} & \textbf{1/2} & 1/1 & - & 0/1 & 0 \\
\shortstack[l]{79451d4e} & \textbf{3/3} & 1/1 & 1/1 & 1/1 & 0 &
\shortstack[l]{d3d51621} & \textbf{1/1} & - & 1/1 & - & 0 \\
\shortstack[l]{487e891e} & \textbf{3/3} & - & 2/2 & 1/1 & 0 &
\shortstack[l]{67462d31} & \textbf{1/2} & - & 1/2 & - & 0 \\
\shortstack[l]{d3b5b821} & \textbf{2/3} & - & 0/1 & 2/2 & 0 &
\shortstack[l]{fb8888cb} & \textbf{2/2} & - & 2/2 & - & 0 \\
\shortstack[l]{415b8e9d} & \textbf{5/5} & - & 3/3 & 2/2 & 0 &
\shortstack[l]{79d6665b} & \textbf{3/3} & - & 3/3 & - & 0 \\
\shortstack[l]{6b14ac4c} & \textbf{3/3} & - & 1/1 & 2/2 & 0 &
\shortstack[l]{54197c4b} & \textbf{3/3} & - & 2/2 & 1/1 & 0 \\
\shortstack[l]{54995d2b} & \textbf{1/1} & - & - & 1/1 & 1 ME &
 & & & & & \\
\midrule
 & & & & & &
 \textbf{Total} & \textbf{93} & \textbf{9} & \textbf{42} & \textbf{42} & 3 (1FP) \\
\bottomrule
\end{tabular}

\end{minipage}
\endgroup
\end{table*}

\para{Bug-reachability results.} 
Table~\ref{tab:rq1-bugs} shows the bug-reachability results of different fuzzers on the \dsa.
\shaohua{Since the bug-reachability results in Table~\ref{tab:rq1-bugs} are stable across the five repetitions: each campaign reaches the same set of bugs in all five runs, the variance and standard deviation of bug counts are all 0. Therefore, we do not show these uninformative data in the table.}
\wanxu{Since \dsa contains protocol-specific vulnerabilities, this experiment supplies the corresponding audit-derived invariants as custom oracles and measures whether each fuzzer can synthesize transactions that reach the audited vulnerable states.}
\wanxu{Under this reachability setting,} \tool demonstrates the strongest bug-reachability capability in all projects. Out of all 109 bugs, \tool successfully detects more than 85\% of them. Notably, \tool detects \emph{all the 9 Critical bugs}, the most concerning vulnerabilities in the real world. On 23 out of 37 projects, \tool finds all the bugs.
In comparison, \ityfuzz only detects 3 bugs, with all of them being \emph{Major} or \emph{Medium}, and no \emph{Critical} bugs.

\noindent
\textbf{Transaction Length.}\wanxu{The generated transactions are multi-call transactions: across the \dsa and \dsb campaigns, a generated transaction contains 5.29 Move function calls on average, and successful \dsa witnesses contain 5.88 calls on average.}
\zq{The length of real-world incidents in \rqfour is much longer, with 7.46 calls on average, and \nemo, for example, requires 10 Move function calls, even for a minimized reproduction.}
\wanxu{Thus, triggering these bugs requires a long process including ensuring type validity, correcting type substitution, and solving numeric path constraints within the same executable transaction.}

\noindent
\textbf{Predefined Oracle Results.}\issta{
We further present the results of the predefined oracles against \movescan on \dsa and \dsb in Table~\ref{tab:rq1-movescan-gap}.
\tool outperforms \movescan in both datasets and covers every finding \movescan reports.
On \dsa and \dsb, \tool covers additional 13 (11 \emph{Precision Loss}, 1 \emph{Type Conversion}, 1 \emph{Unnecessary Bool Judgement}) and 28 (all \emph{Precision Loss}) bugs, respectively.
While \movescan reports 8 and 12 false positive bugs on \dsa and \dsb, \tool has no false positive reports.
We attribute such advantages to two reasons: (1) \movescan scans these vulnerabilities within a single function context statically, while \tool is capable of tracking values and invariants across multiple call boundaries at runtime; and (2) the concolic execution of \tool can provide the precise context for the predefined oracles. Note that all the bugs in Table~\ref{tab:rq1-bugs} cannot be detected by \movescan because the oracles of \movescan are only capable of detecting simple Move smart contract defects, while real-world vulnerabilities are usually complicated and bound to specific business models.}

\para{\suifuzzer Deficiency.} 
We do not include the result of \suifuzzer in Table~\ref{tab:rq1-bugs} because it does not yield any findings even after we fix many bugs in it.
\suifuzzer always aborts early in our experiments because it does not support shared objects. We then investigate that even if \suifuzzer could support shared objects, it would still fail all the fuzzing experiments because it still has several inherent design flaws: (1) it will forge objects of any types by mutating the objects unconditionally, leading to massive amount of false positives; (2) it only supports testing a single Move smart contract with selected functions while the real-world defects mostly span multiple ones; (3) it only supports filling up objects from the existing global owned objects, making it unable to correctly handle ``Hot Potato'' objects; and (4) it does not consider types of objects \rev{and usually fills in objects with unmatched types}.

\begin{figure}
    \begin{minted}[xleftmargin=1em,numbersep=3pt,fontsize=\footnotesize,linenos,escapeinside=@@]{rust}
public fun mul_div(a: u128, b: u128, c:u128): u128 { a * b / c @\label{line:typloss}@ }
public fun compute_boost_factor(vestar_amount: u128, farm_amount: u128, total: u128): u64 {
    // precision loss because division before multiplication
    let dividend = mul_div(vestar_amount, FACTOR * 3, total) * FACTOR; @\label{line:loss1}@
    let divisor = mul_div(farm_amount, FACTOR * 2, total);
    ((dividend / divisor) + FACTOR) / (FACTOR / 100)
}
public fun calculate_boost_weight(amount: u128, boost_factor: u64): u128 {
    // precision loss because division before multiplication
    amount * (boost_factor as u128) / BOOST_FACTOR_PRECESION @\label{line:loss2}@
}
public fun update_boost_factor(...) {
    let boost_factor = compute_boost_factor(...);
    let weight = calculate_boost_weight(asset_amount, boost_factor);
}
    \end{minted}
    \caption{A \textit{Precision Loss} defect of the \textit{starswap-core-aptos} project in \dsb.}\label{lst:precision-loss}
\end{figure}

\begin{table}[t]
\begingroup
\scriptsize
\centering
\setlength{\abovecaptionskip}{3pt}
\setlength{\belowcaptionskip}{0pt}
\caption{Results of the predefined oracles on \dsa and \dsb. The projects where \tool and \movescan report the same results are abbreviated.}
\label{tab:rq1-movescan-gap}
\setlength{\tabcolsep}{6pt}%
\renewcommand{\arraystretch}{1.0}
\rowcolors{2}{lightgray!50}{}
\begin{tabular}{lcc|lcc}
\toprule
\textbf{\dsa Projects} & \textbf{\tool} & \textbf{\movescan} & \textbf{\dsb Projects} & \textbf{\tool} & \textbf{\movescan} \\
\midrule
8bc50340 & 18 & 26 (8 FP) & utils & 0 & 6 (6 FP) \\
73c07115 & 1 & 0 & originmate & 39 & 16 \\
e3564d47 & 29 & 27 & pyth-crosschain-sui & 2 & 8 (6 FP) \\
074ef533 & 2 & 0 & Starswap-aptos farming & 8 & 4 \\
8f0c3e4c & 1 & 0 & LiquidswapRouterV2 & 1 & 0 \\
fb8888cb & 8 & 1 &  &  &  \\
\midrule
\textbf{Total} & \textbf{816} & \textbf{811 (8 FP)} & \textbf{Total} & \textbf{146} & \textbf{130 (12 FP)} \\
\bottomrule
\end{tabular}
\endgroup
\end{table}

\para{Case Study in \dsb.}
We demonstrate our advantage over \movescan with two \textit{Precision Loss} vulnerabilities from the same project from the \textit{starswap-core-aptos} project of \dsb on the Aptos chain.
The first \textit{Precision Loss} happens at Line~\ref{line:loss1} in Figure~\ref{lst:precision-loss}. The root cause is that the division of \verb|total| happens before the multiplication of \verb|FACTOR|.
The second \textit{Precision Loss} is at Line~\ref{line:loss2} in Figure~\ref{lst:precision-loss}, where the division of \verb|FACTOR| to \verb|boost_factor()| happens in the function \verb|compute_boost_factor()| before the multiplication of \verb|amount| in \verb|calculate_boost_weight()| when calling the function \verb|update_boost_factor()|.
\movescan misses both vulnerabilities because it only detects within a single function call context and does not trace values across function boundaries.
In addition, it only supports a very limited fixed pattern like \verb|(a / b) * c| with exactly three operands.
\tool finds both \textit{Precision Loss} vulnerabilities because we track symbolic expressions for each value across functions and accurately identify potential precision loss.

\noindent
\textbf{Time-to-Reach.} \wanxu{We further analyze the time to the first oracle violation for the 93 \dsa bugs reached by \tool. The median time to bug is 30.7 seconds, and the mean is 41.5 seconds. Overall, 91.5\% of successful detections occur within one minute and 99.1\% occur within five minutes. Aggregating each bug by its median time across repeated runs gives the same trend: 89 of 93 bugs (95.7\%) have a median time to bug below one minute, and all 93 bugs have a median below five minutes. By severity, the per-bug median time is 40.7 seconds for \textit{Critical}, 36.0 seconds for \textit{Major}, and 29.3 seconds for \textit{Medium} bugs.}

\para{False negatives in \tool.} While \tool shows strong capability of finding bugs on \dsa, we still miss 15\% (16) bugs. \wanxu{We classify them into two categories.}
\wanxu{First, 7 bugs are out of scope for our current oracle and threat model.} For instance, some contracts are vulnerable to Denial-of-Service attacks due to iterating over too large vectors, while others are vulnerable to incorrect logging during execution. \wanxu{As a smart contract fuzzer, \tool cannot detect vulnerabilities that do not affect executable integrity.}
\wanxu{Second, 9 bugs require administrator privileges to trigger.} For example, centralization risk is a common vulnerability where the administrator has too much power. Both \tool and the baselines \ityfuzz and \suifuzzer do not assume any special capabilities for testing.
\wanxu{If these 16 out-of-scope or privileged-only bugs are excluded, \tool reaches all remaining \dsa ground-truth bugs.}
\wanxu{A representative privileged-only miss is the centralization-risk finding in project \textit{916a}. Three in-game currency modules each define a fungible token, create the corresponding \texttt{TreasuryCap} during initialization, and transfer that cap to the deployment sender. Their public \texttt{mint\_and\_transfer} functions then accept the corresponding mutable \texttt{TreasuryCap}, mint an attacker-chosen amount, and transfer the resulting tokens to an arbitrary recipient. This behavior is exactly the audited centralization risk, but exercising it requires the administrator-owned \texttt{TreasuryCap} while \tool does not assume access to such privileged objects.}
We will discuss more limitations in Section~\ref{sec:discussion}.

\begin{figure*}[t]
    \centering
    \includegraphics[width=1.0\linewidth]{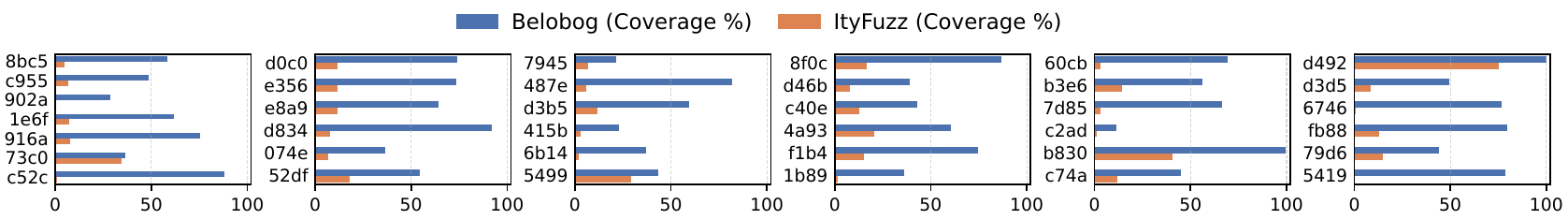}
    \caption{Edge coverage of \tool and \ityfuzz on the projects from \dsa.}
    \label{fig:coverage}
\end{figure*}

\subsection{RQ2: Coverage and Throughput}\label{sec:coverage}
We collect the edge coverage of each project during fuzzing. Figure~\ref{fig:coverage} shows the average results in each project.
Across all 36 out of 37 projects in \dsa, \tool reaches significantly higher coverage than \ityfuzz on all of them.
Specifically, \tool achieves an average coverage of 55.6\%, whereas \ityfuzz achieves only 8.8\%.
\wanxu{The coverage results are also stable across repetitions: a run-to-run significance test on the five coverage runs gives $p=0.889$, indicating no statistically significant difference among repeated runs.}
\issta{\ityfuzz reaches similar code coverage to \tool only on project 73c0. \issta{The reason is that the project needs administrator privileges for deeper testing, which neither \tool nor \ityfuzz assumes.}}
\issta{For the fuzzing throughput, \tool is ten times faster than \ityfuzz. The reason is that \ityfuzz is type-unaware, leading to lots of computation waste in enumerating objects and calls to consume the hot potatoes.}\wanxu{The concolic executor remains lightweight in these campaigns. Across all runs, \tool invokes the solver 2,193,310 times, with an average solver time of 21.86~ms per invocation and a timeout rate of only 0.08\%. Solver time accounts for 3\% of the total fuzzing time, indicating that the 500~ms per-transaction bound keeps constraint solving from dominating the fuzzing loop.}

\begin{table}[t]
\begingroup
\scriptsize
\centering
\setlength{\abovecaptionskip}{3pt}
\setlength{\belowcaptionskip}{0pt}
\caption{Bug-reachability results of \tool and its variants on \dsa.}
\label{tab:ablation}
{\setlength{\tabcolsep}{2pt}%
\renewcommand{\arraystretch}{1.05}%
\rowcolors{2}{lightgray!50}{}
\begin{tabular}{lccccccc}
\toprule
\textbf{Category} & \textbf{Ground Truth} & \textbf{\tool} & \textbf{\noconcolic} & \textbf{\notg} & \textbf{\noty} & \textbf{\noall} \\
\midrule
Critical           & 9  & 9  & 5   & 0   & 4   & 0   \\
Major              & 53 & 42 & 28  & 2   & 26  & 1   \\
Medium             & 47 & 42 & 33  & 1   & 28  & 1   \\
\midrule
Total    & 109 & 93 & 66 & 3 & 58 & 2 \\
\bottomrule
\end{tabular}}
\endgroup
\end{table}

\subsection{RQ3: Ablation Study}
In order to understand the impact of individual components in \tool, we additionally implement four variants by excluding core components and evaluate their bug-reachability capabilities on \dsa.
\begin{itemize}[leftmargin=20pt]
    \item \noconcolic: In this variant, we remove the concolic executor from \tool.
    \item \notg: In this variant, we no longer generate the \wanxu{dependency graph} but instead use a random havoc mutator as implemented in \ityfuzz.
    \item \noty: In this variant, we do not fill type parameters and skip functions requiring type parameters.
    \item \noall: In this variant, we remove concolic execution, \wanxu{dependency graphs}, and type parameter support.
\end{itemize}


The last column in Table~\ref{tab:ablation} shows the result of removing all three key components in \tool. Only two bugs can be detected, suggesting the criticality of the three components. Below, we discuss the results of each variant.

\smallskip
\noindent\textbf{Impact of \wanxu{dependency graph}.} 
The \notg column in Table~\ref{tab:ablation} shows the bug-reachability result after removing the \wanxu{dependency graph}.
Compared to our default \tool, \emph{none of the Critical} bugs can be detected, and the overall bug detection ratio drops from 85\% to nearly zero.
The main reason is that without \wanxu{dependency graphs}, \notg can no longer guarantee that ``Hot Potato'' objects are consumed properly, and thus many functions can not be covered.

\smallskip
\noindent\textbf{Impact of type parameters.} 
The \noty column in Table~\ref{tab:ablation} shows the bug-reachability result after removing the support of type parameter.
Compared to our default \tool, only 4 \emph{Critical} bugs can be detected, and the overall bug detection ratio drops from 85\% to 53\%.
The reason for the bad performance is that 37\% of the functions in \dsa have at least one type parameter and thus all transactions generated by \noty can not make a valid call to them. 

\smallskip
\noindent\textbf{Impact of concolic executor.} 
The \noconcolic column in Table~\ref{tab:ablation} shows the bug-reachability result after removing the concolic executor.
Compared to our default \tool, only 5 \emph{Critical} bugs can be detected, and the overall bug detection ratio drops from 85\% to 60\%.
This suggests that concolic execution is crucial to exploring critical paths in smart contracts.

\begin{figure}
    \begin{minted}[xleftmargin=1em,numbersep=3pt,fontsize=\footnotesize,linenos,escapeinside=@@]{rust}
fn checked_shlw(v: u256): u256 { /// private function for implementing left shift overflow check
    if (v > (0xFFFFFFFFFFFFFFFF << 192)) { (0, true) } @\label{line:leftshift}@ // incorrect check with too big value.
    else { (arg0 << 64, false) }
}
fn get_delta(liquidity: u128, price_low: u128, price_high: u128): u64 { /// private function
    let price_delta: u128 = price_high - price_low;
    let prod: u256 = liquidity * price_delta;
    let (shifted, overflow) = checked_shlw(prod);
    if (overflow) { abort; } // abort if a left shift overflow is detected
    (shifted / (price_low * price_high)) as u64 // MoveVM aborts for u128 values exceeding u64.
}
public struct AddLiquidityReceipt<T> { repay_amount: u64 } /// "Hot Potato" type 
public fn add_liquidity<T>(amount: u64): AddLiquidityReceipt<T> {
    let (price_low, price_high) = ...;
    let repay_amount = get_delta(amount, price_low, price_high); @\label{line:repaya_amount}@
    AddLiquidityReceipt { repay_amount }
}
public fn repay_liquidity<T>(coin: Coin<T>, receipt: AddLiquidityReceipt<T>) { 
    let { repay_amount } = receipt;
    assert!(coin.value == repay_amount); @\label{line:assert_coin_value}@
}
    \end{minted}
    \vspace{-10pt}
    \caption{The simplified smart contract from the buggy \cetus.}\label{lst:cetus}
\end{figure}

\subsection{RQ4: On-Chain Incidents Reproduction}\label{sec:eval-realworld}

\issta{We select two recent catastrophic on-chain incidents on the Sui blockchain, known as \cetus\cite{cetusreport} and \nemo\cite{nemoreport}, which resulted in more than \$200 million and \$2.6 million financial losses, respectively. \tool is capable of reproducing the full exploits of both incidents with just the \emph{Earning Profits} oracle. We discuss \cetus in this section.}

Figure ~\ref{lst:cetus} presents a simplified \cetus smart contract.
Although most arithmetic overflows are properly checked by the MoveVM, left shift is not.
Thus, there is a left shift check function \verb|checked_shlw()| implemented in the \cetus to prevent such overflows.
However, the function incorrectly encodes the left shift overflow threshold as \verb|v > (0xFFFFFFFFFFFFFFFF << 192)| (Line~\ref{line:leftshift}), which is much bigger than it is supposed to be, \ie, \verb|v >= (1 << 192)| as the later patch shows.
Therefore, any values between the two boundaries, \ie, \verb|(1 << 192) < v <=| \verb|(0xFFFFFFFFFFFFFFFF << 192)|,  will lead to an overflowed value being returned.

The real-world exploit constructed a transaction with the call sequence
$$\texttt{add\_liquidity(AMOUNT)} \Rightarrow \texttt{repay\_liquidity(...)}$$
with a carefully constructed value \verb|AMOUNT|. The \verb|add_liquidity()| calls \verb|get_delta()| to calculate the tokens (money) that need to be paid back later, and \verb|get_delta()| calls \verb|checked_shlw()| to avoid too large values.
However, a carefully crafted large value \verb|AMOUNT| can escape the overflow check in \verb|checked_shlw()|, leading to a much smaller \verb|repay_amount| in line~\ref{line:repaya_amount}.
In other words, the transaction sender can drain a considerable amount of assets with little cost.

\para{Fuzzing \cetus with \tool.} 
As shown in Figure~\ref {lst:cetus}, the two functions, \verb|add_liquidity()| and \verb|repay_liquidity()|, have to respectively produce and consume a ``Hot Potato'' object that has the type \verb|AddLiquidityReceipt<T>|.
Therefore, using our \wanxu{dependency graph}, \tool can generate the expected function-call sequence.
However, the \wanxu{dependency graph} itself alone is not enough to trigger this vulnerability for two reasons: (1) triggering the overflow within \verb|checked_shlw()| is not trivial because the last line of \verb|get_delta()| casts the value from \verb|u256| to \verb|u64|, resulting in a much tighter bound. 
Our calculation reveals that the chance of randomly mutating a value to trigger a valid overflow is as small as $2^{-30}$; 
(2) line~\ref{line:assert_coin_value} in \verb|repay_liquidity()| enforces that the input \verb|coin| has to be equal to the \verb|repay_amount| calculated in line~\ref{line:repaya_amount}.
These constraints make it extremely challenging to generate a successful exploit.
Fortunately, \tool can readily address both obstacles using its concolic executor.

When fuzzing starts, \tool forks the Sui blockchain at checkpoint 148114817, right before the incident happens. Then, it initializes the on-chain object pool with available shared objects obtained from the forked chain.
\tool is capable of detecting the \textit{SHL overflow} issue in a few seconds and reproducing the whole exploit with the \textit{Earning Profits} oracle in less than 3 hours, where most of the time is spent on executing seeds.

\subsection{RQ5: Real-world Application}

\issta{We further apply \tool on another three projects that are still under auditing. \tool can exploit one critical vulnerability leading to draining pool rewards by \emph{Earning Profits} and one medium vulnerability by \emph{Precision Loss} \textit{without manually crafting oracles}. By further implementing oracles for the projects, \tool additionally reports 1 critical, 2 major, and 2 medium vulnerabilities. All these findings have been confirmed by the project developers and fixed before deployment.}

\begin{figure*}[t]
    \centering
\begin{minted}[xleftmargin=1em,numbersep=3pt,fontsize=\footnotesize,linenos,escapeinside=@@]{rust}
public fun swap<X, Y>(...) { pool_fee_growth += fee_amount; @\label{line:fee_growth}@ }
public fun open_position<X, Y>(...): Position {
    Position { liquidity: 0, fee_growth_last: 0, rewards: 0, }
}
public fun add_liquidity<X, Y>(position, liquidity, ...) {  position.update(liquidity); }
/// internal helper
fun update(position, delta, ...) {
    position.liquidity += delta; // Bug: should have updated position.liquidity in the end @\label{line:update_bug}@
    // rewards are non-empty for newly created positions because position.liquidity is not zero
    let rewards = (pool_fee_growth - position.fee_growth_last) * position.liquidity; @\label{line:distribute_fee}@
    position.rewards += rewards; // position.rewards should be zero for newly created positions
    position.fee_growth_last = pool_fee_growth;
    // Fix: update position.liquidity here
}
public fun collect_fees(position): Coin { position.rewards @\label{line:colllect_reward}@ }
\end{minted}
    \caption{A simplified code snippet of a real-world bug that allows anyone to drain pool rewards.}
    \label{fig:case-hop}
    \vspace{-10pt}
\end{figure*}

\para{\issta{Real-world Bug Incident Study.}}
The simplified code of the target project, as illustrated in the Figure~\ref{fig:case-hop}, is a Decentralized Exchange (DEX) that accumulates fees into \verb|pool_fee_growth| (Line~\ref{line:fee_growth}) that are later distributed to the positions of liquidity providers (Line~\ref{line:distribute_fee}) as rewards.

However, the code that updates position liquidity has a serious flaw: it calculates rewards based on the \textit{updated} position liquidity as Line~\ref{line:update_bug} suggests, which causes a newly created position to receive non-zero rewards from \verb|position.rewards|. This allows any user to withdraw all previous rewards by creating a position with the minimum liquidity requirement using the sequence

$$\texttt{open\_position()} \Rightarrow \texttt{add\_liquidity(position, 1)} \Rightarrow \texttt{collect\_fees(position)}$$

\para{\issta{Fuzzing with \tool}}
\tool will compile the target package and deploy it on a locally forked blockchain environment. \tool can generate the full exploit sequence in less than 30 minutes because it triggers the \emph{Earning Profits} oracle by withdrawing the rewards of previous swaps.

\section{Discussion}\label{sec:discussion}

\noindent
\textbf{Automated Testing Smart Contracts.}
Both \cetus and \nemo studied in our evaluation are heavily audited by the leading Web3 auditing companies, but still, they fail to prevent these critical vulnerabilities.
This suggests that humans are prone to making errors, and pure manual efforts are not enough to protect the security of smart contracts.
\issta{With \tool, these losses could have been avoided.}
Therefore, continuous and automated testing for the smart contracts is essential to secure deployed smart contracts.
While there are already lots of automated testing tools on the prevailing Ethereum Virtual Machine (EVM), we find that the Move ecosystem greatly lacks such tools and frameworks.
Therefore, we hope that \tool can shed light on the Move smart contracts testing by providing an extensible and general fuzz testing framework.

\smallskip
\noindent
\textbf{Oracle Automation.} \wanxu{While \tool has demonstrated its effectiveness in reaching audited vulnerable states and achieved high coverage and throughput, one of the limitations is that \tool needs extra manual efforts to build oracles when testing real-world smart contracts that have a custom business model.}
\wanxu{We work around this by providing predefined oracles for common defects and allowing users to write protocol-specific oracles in Move language instead of modifying the fuzzer itself.}
It is an interesting direction to explore automated fuzzing oracle synthesis for Move smart contracts, given their strongly typed nature.
It is possible to apply some existing approaches~\cite{actlifter,defiranger,defort,sleuth} to extract high-level smart contract semantics to help build oracles automatically.
We deem this as one of the interesting directions for our future research.

\smallskip
\noindent
\textbf{State Dependency.} \wanxu{Although \tool builds a dependency graph to model type dependencies among functions, it does not model arbitrary inner state dependency.}
\wanxu{Our evaluation suggests that our concolic executor can solve direct dependencies bound to input parameters, as the \cetus study in Section~\ref{sec:eval-realworld} shows.}
\wanxu{However, \tool may miss bugs that require creating a hidden state through a long history of transactions before the vulnerable transaction becomes executable, especially when that state is not reflected in function signatures or object abilities.}
\wanxu{In our analysis of false negatives in \dsa, we did not observe false negatives caused by such hidden state dependencies.}
It is also possible to integrate dataflow analysis to understand the state dependency, like Smartian\cite{smartian}.
We leave this as our future work.

\section{Related Work}

\noindent
\textbf{Move Language Testing.} So far, there are a limited number of tools to test Move contracts automatically.
Move Prover~\cite{moveprover} is a pioneer in translating the Move language to its custom intermediate verification language to verify pre- and post-conditions.
However, it requires manual effort to write specifications and only supports a limited set of oracles, making it less practical given the fancy business models of the DeFi ecosystem and learning costs~\cite{movescan}.
MoveLint~\cite{movelint} is an open-source tool that detects several Move defects by analyzing the Abstract Syntax Tree of smart contracts. While it supports a wide range of oracles, it suffers from false positives due to the inaccuracy of static rules. \movescan~\cite{movescan} instead translates the Move bytecode into stackless intermediate representation and generates a Control Flow Graph to facilitate code analysis with extended oracle definitions.
\ityfuzz~\cite{ityfuzz} is originally designed for fuzzing bytecode on Ethereum Virtual Machine (EVM), it also ports its snapshot-based fuzzing approaches to MoveVM afterwards.
\suifuzzer\cite{suifuzzer} is an coverage-guided fuzzer supporting both stateful and stateless fuzzing with custom properties.

\smallskip
\noindent
\textbf{Smart Contract Fuzzing.} Fuzzing is an efficient way to test smart contracts.
sFuzz~\cite{sfuzz} uses strategies adapted from AFL to improve the code coverage.
Smartian~\cite{smartian} proposes both static and dynamic dataflow analysis to solve the complex data dependencies. 
ConFuzzius~\cite{confuzzius} hybridizes symbolic execution, data dependency analysis, and taint analysis to exercise deeper bugs.
VULSEYE~\cite{vulseye} novelly models the smart contract fuzzing as stateful directed fuzzing and guides the fuzzing by vulnerabilities.
VERITE~\cite{verite} builds a first profit-centric fuzzing framework with a gradient descent method to maximize the exploits.
LLM4Fuzz~\cite{llm4fuzz} attempts to utilize the capability of Large Language Model (LLM) to guide the smart contract fuzzing.
However, all of them focus on smart contracts on EVM and do not consider various type constraints in smart contracts. Thus, it is non-trivial to port them to support Move smart contracts as we have tried with \ityfuzz.
\issta{Meanwhile, it would also be hard to port our methods to Solidity and EVM because of its hashing table style storage, where type information is much less visible from the public function signatures. For instance, the interface of standard ERC20~\cite{erc20} that represents assets on EVM uses primitive values like \textit{address} and \textit{uint256} while Move has strongly-typed \textit{Coin<T>} for assets. Therefore, the benefits of building \wanxu{dependency graphs} for public interfaces on Solidity and EVM are very limited.}

\smallskip
\noindent
\textbf{API Fuzzing.} Generating a sequence of function calls to test the smart contract is similar to the traditional API fuzzing for libraries.
GraphFuzz~\cite{graphfuzzing} models the execution trace as a dataflow graph to test the low-level library APIs.
Minerva~\cite{minerva} utilizes dynamic mod-ref analysis to test the browser APIs.
APICraft~\cite{apicraft} leverages both static and dynamic information to build a fuzzer driver for testing closed-source SDKs.
OGHARN~\cite{noharness} proposes the novel oracle-guided harnessing to automatically fuzz APIs.
\issta{There are also several works utilizing type information to test code, like RPG~\cite{rpg}, RuMono~\cite{rumono}, TypeNFuzz~\cite{typenfuzz}, and SYPET~\cite{sypet}. However, the Move Language type safety rules, including the ``Hot Potato'' mechanism, present additional challenges and gaps. Therefore, applying such methods can not work as it is, though it is interesting to explore how these insights could help enhance the capabilities of \tool.}
\section{Conclusion}

In this paper, we present \tool, the first effective fuzzing framework for Move smart contracts.
We identified three key challenges in generating valid transactions for Move smart contracts, \ie, type safety of objects, type parameters of functions, and type abilities of objects.
To address these challenges, we propose a \emph{\wanxu{dependency graph}}-guided fuzzing approach, in which we use a \wanxu{dependency graph} to model all type constraints of the target smart contract. When fuzzing starts, \tool will query the \wanxu{dependency graph} to generate or mutate transactions that satisfy all the type requirements. 
To further tackle the strict checks that are common in smart contracts, we design and implement a concolic executor in \tool.
Our extensive evaluation on two real-world datasets demonstrates the capability of \tool.

\section*{Acknowledgment}

This research is supported by the National Research Foundation, Singapore, and DSO National Laboratories under the AI Singapore Programme (AISG Award No: AISG4-GC-2023-008-1B); by the National Research Foundation Singapore and the Cyber Security Agency under the National Cybersecurity R\&D Programme (NCRP25-P04-TAICeN); and this research is part of the IN-CYPHER programme and is supported by the National Research Foundation, Prime Minister’s Office, Singapore under its Campus for Research Excellence and Technological Enterprise (CREATE) programme.. Any opinions, findings and conclusions, or recommendations expressed in these materials are those of the author(s) and do not reflect the views of the National Research Foundation, Singapore, Cyber Security Agency of Singapore, Singapore.

\section*{Data Availability}

Our tool is available at ~\cite{artifact}.

\bibliographystyle{ACM-Reference-Format}
\bibliography{acmart}










\end{document}